\documentclass[a4paper,fleqn,usenatbib]{mnras}

\usepackage[T1]{fontenc}
\usepackage{ae,aecompl}

\usepackage{graphicx}	
\usepackage{amsmath}	
\usepackage{amssymb}	

\newcommand{\Msun}{M_{\odot}}
\newcommand{\Rsun}{R_{\odot}}

\newcommand{\Mwd}{M_{\mathrm{WD}}}

\title[CAML can solve major problems of CV evolution]{Three in one go: consequential angular 
momentum loss can solve major problems of CV evolution}

\author[Schreiber, Zorotovic \& Wijnen]{
M.R. Schreiber,$^{1}$\thanks{matthias@ifa.uv.cl}
M. Zorotovic,$^{1}$
and T.P.G. Wijnen$^{1,2}$
\\
$^{1}$Institute of Physics 
and Astronomy, Universidad de Valparaiso, 
Av. Gran
Bretana 1111, Valparaiso, Chile\\
$^{2}$Department of Astrophysics/IMAPP, Radbound University Nijmegen, PO box
9010, NL-6500 GL Nijmegen, The Netherlands\\
}

\date{Accepted XXX. Received YYY; in original form ZZZ}

\pubyear{2015}

\begin{document}
\label{firstpage}
\pagerange{\pageref{firstpage}--\pageref{lastpage}}
\maketitle

\begin{abstract}
The average white dwarf (WD) masses in cataclysmic variables (CVs) have been measured to significantly
exceed those of single WDs, which is the 
opposite of what is theoretically expected. 
We present the results of binary population synthesis models 
taking into account consequential angular 
momentum loss (CAML) that is assumed to increase 
with decreasing WD mass. This approach can 
not only solve the WD mass problem, but also brings 
in agreement theoretical
predictions and
observations of the orbital period distribution and 
the space density of CVs. We speculate
that frictional angular 
momentum loss following nova eruptions might 
cause such CAML and could thus be the missing ingredient 
of CV evolution.  
\end{abstract}




\section{Introduction}\label{s:intro}
The first theoretical binary population 
synthesis (BPS) models of Cataclysmic 
Variables (CVs) revealed dramatic disagreements 
between observations 
and theoretical predictions:
1.\, CVs below the period gap 
should make up $\sim$\,99 per cent of the population
\citep{kolb93-1}, 
while roughly the same number of systems is 
observed 
below and above the gap \citep{knigge06-1}.  
2.\, The observed orbital period minimum is at 
76\,min \citep{knigge06-1}, 
while the predicted orbital period minimum 
is 65-70\,min \citep{kolb+baraffe99-1}. 
3.\, A strong accumulation of systems at the orbital 
period minimum is predicted, but not observed 
\citep{patterson98-1,knigge06-1}. 
4.\, The predicted space density \citep[e.g.][]{ritter+burkert86-1,kolb93-1} exceeds the observed 
one by 1-2 orders of magnitude
\citep[e.g.][]{patterson98-1,pretorius+knigge12-1}. 

Within the last decade, two of these long standing 
problems of CV evolution seem to have been solved.  
First, the problem of the missing spike at the 
orbital period minimum was solved as 
recent deep surveys identified a previously 
hidden population of faint CVs \citep{gaensickeetal09-2}. 
Secondly, the orbital period minimum problem 
can be solved if additional angular 
momentum loss, apart from gravitational radiation, is 
assumed to be present in systems with fully 
convective donor stars \citep{kniggeetal11-1}. 

However, recent observational advances 
not only solved previous problems but also 
generated a new one, which is currently 
perhaps the most severe.
BPS models 
predict an average white dwarf (WD) 
mass of $\lesssim0.6\Msun$
\citep[e.g.][]{politano96-1} and large numbers of 
CVs containing helium core (He-core) WDs. 
Yet, the observed average
mass is $\sim0.83\Msun$ and not a single system 
with a definite He-core WD primary 
has been identified so far 
\citep[][hereafter ZSG11]{zorotovicetal11-1}. 
As shown by ZSG11, this absence of CVs with low-mass primaries 
can no longer be explained as an observational 
bias as suggested earlier 
by \citet{ritter+burkert86-1}.
Moreover, the problem is also clearly not related 
to our limited 
understanding of common envelope (CE) evolution 
as speculated recently by \citet{geetal15-1}.  
This is because low-mass WDs are found in large
numbers in samples of detached post 
common envelope binaries 
\citep{rebassa-mansergasetal11-1} 
and the observed WD mass distribution of 
these objects is in good agreement 
with the model
predictions \citep{toonen+nelemans13-1,zorotovicetal14-1,camachoetal14-1}.  
Finally, in \citet{wijnenetal15-1} we have 
shown recently that the two
scenarios suggested by 
ZSG11, i.e. mass growth in CVs or 
a preceding phase of thermal time scale 
mass transfer 
for many CVs cannot solve the WD mass 
problem in CVs either. 

Here we present BPS models of CVs 
taking into account angular momentum loss 
generated by the mass transfer in CVs.  
This consequential angular momentum 
loss (CAML) 
can drive CVs with low-mass WDs
into dynamically unstable mass transfer
which solves the WD mass problem. 
Interestingly, this approach simultaneously 
solves the space density problem and 
brings into agreement the predicted and observed 
orbital period distributions. 

\section{General model assumptions}\label{s:model}
The aim of this work is to evaluate whether 
assuming CAML can explain the observed 
WD masses in CVs. To that end we
simulate CV evolution with 
different CAML approaches combined 
with standard assumptions for CV evolution.  

We generate a population of $10^7$ 
initial main sequence (MS) binaries. 
The star 
masses ($M_1, M_2$) are distributed 
according to the 
initial-mass function of \citet{kroupaetal93-1} 
for the primary and a flat initial mass-ratio 
distribution \citep{sanaetal09-1}. 
$M_1$ ranges from $0.8$ 
to $9\Msun$ and $M_2$
from $0.05$ to $9\Msun$ (requesting of course $M_2<M_1$).
The initial orbital separation is 
assumed to be flat in $\log\,a$ 
\citep{popovaetal82-1, kouwenhovenetal09-1},
ranging from $3$ to $10^4\Rsun$. 

The sample of generated MS binary 
stars is then evolved until the end of the 
common envelope (CE)
phase using the binary-star evolution (BSE) 
code from \citet{hurleyetal02-1} 
assuming a CE efficiency of $\alpha=0.25$ 
\citep{zorotovicetal10-1}.  
The generated population of zero age 
post-CE binaries 
is subsequently evolved to obtain the 
current population of CVs
following the approach used in 
\citet{wijnenetal15-1}. 
We simply assume the Mass-Radius relation for 
CV secondary stars and systemic angular
momentum loss prescriptions as 
derived from observational 
constraints by \citet{kniggeetal11-1}. 
We adjust the scaling factors for 
angular momentum loss due to gravitational radiation and magnetic braking 
to get mass transfer rates in agreement with those
derived from observations \citep{townsley+gaensicke09-1}.
When stable post-CE mass transfer starts   
we smoothly increase the radius of the
secondary star from the MS radius 
to the larger radius of a CV secondary 
\citep[see also e.g.][]{davisetal08-1}.
If the mass of the secondary star drops 
below $0.05\Msun$ we discard 
the system from the sample, 
because the M-R relation is not reliable 
for lower masses.
We furthermore assume that the WD expels 
the accreted mass in repeated nova eruptions 
and thus treat it as constant during 
CV evolution. 
We assume $10$\,Gyrs 
for the age of the Galaxy 
and a constant star formation rate. 
Finally, we ignore CVs that may descend 
from thermal time scale mass transfer, i.e we 
assume these systems make up at most a few per cent 
of the current CV population which is probably correct 
\citep{gaensickeetal03-1}.

While all the above model assumptions are 
quite uncertain, we did run several tests 
varying the uncertain parameters and find 
that the conclusions of this paper are not 
affected. 

\section{Stability of mass transfer and CAML}

The mass ratio 
($M_2/M_{\mathrm{WD}}$) required for stability 
against dynamical time scale mass transfer 
is a crucial part of BPS models
because it separates CVs from systems 
that evolve through a second CE and most likely merge. 
The dividing line can be obtained by equating
the adiabatic mass radius exponent and the mass 
radius exponent of the
secondaries Roche-lobe:  
\begin{equation}\label{eq-f}
\zeta_{ad}\equiv\frac{dln(R_2)}{dlnM_2}_{ad}=\frac{dln(R_{L2})}{dlnM_2}\equiv\zeta_{R_{L2}}. 
\end{equation}
For deeply convective secondary stars
($M_2\lesssim\,0.5\Msun$) polytropic 
models represent a reasonable 
approximation, i.e. $\zeta_{ad}=-1/3$. 
For more massive stars, the size of the 
convective envelope decreases and 
$\zeta_{ad}$ increases steeply
\citep{hjellming89-1}. 
The right hand side of Eq.(\ref{eq-f})
sensitively depends on the assumed CAML.  
We therefore performed BPS 
models with three versions of CAML. 

\subsection{The fully conservative case}

We first assume
conservation of both angular momentum 
and the total mass of 
the binary, i.e. we assume only systemic angular 
momentum loss through
magnetic braking and gravitational radiation but 
no CAML. This assumption is inconsistent 
with mass loss due to nova eruptions, 
which is known
to be of the same order as the mass accreted 
by the WD between two eruptions, but the fully 
conservative model has been previously
used in BPS models 
and we therefore use it as a reference model 
(despite the obvious inconsistency).   

\begin{figure*}
\includegraphics[width=0.73\textwidth]{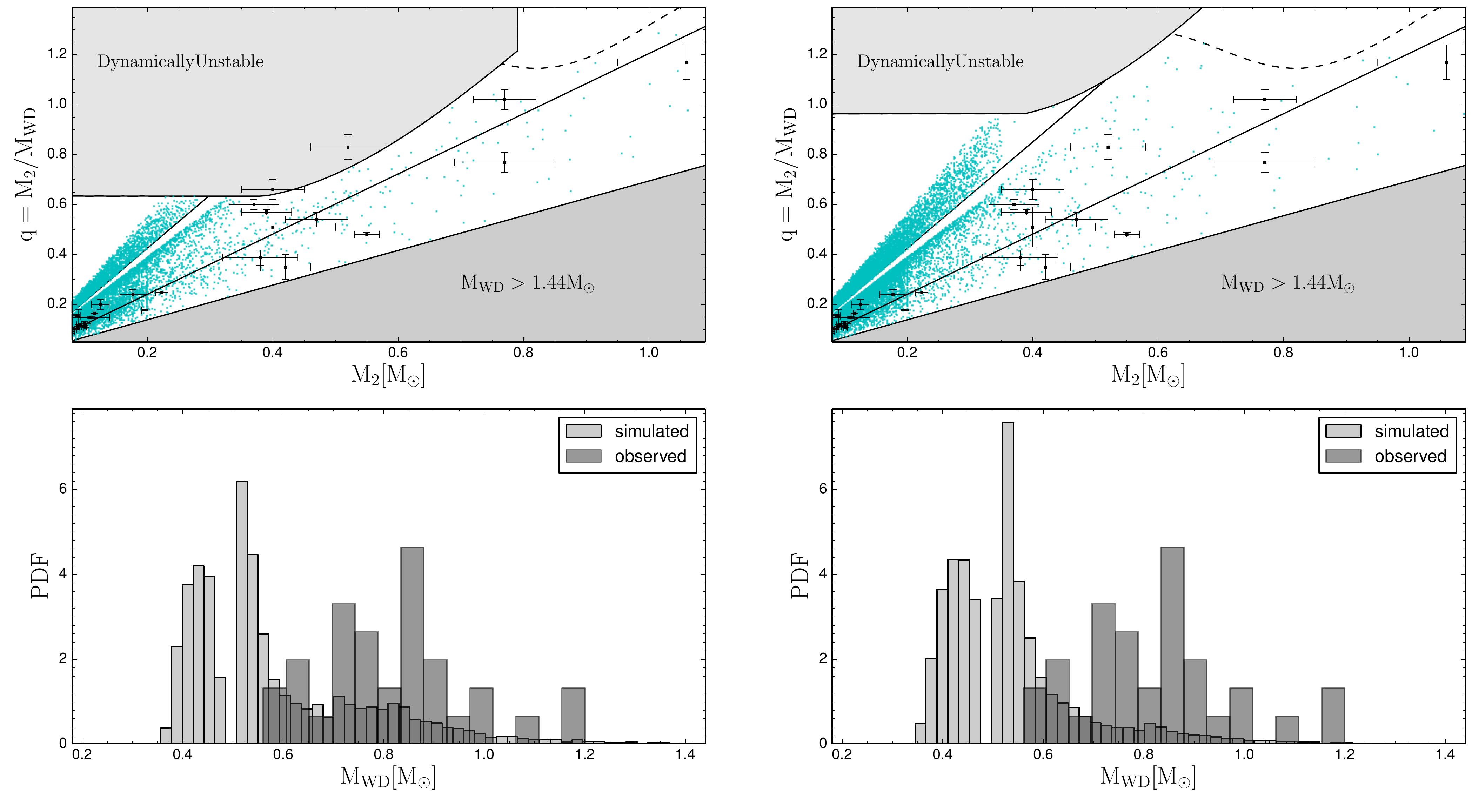}
\caption{\textit{Top}: Observed 
(black squares) and predicted (cyan dots)
CV populations in the q versus $M_2$ diagram 
for the fully conservative case (\textit{left}) 
and for the classical non-conservative model 
(\textit{right}). 
The gray shaded areas represent the forbidden regions either due to 
dynamically unstable mass
transfer or because the WD mass exceeds the Chandrasekhar limit. 
The black solid lines represent the average of the WD masses measured in CVs 
($\Mwd=0.83\Msun$) and the mass limit for He-core WDs ($0.47\Msun$). 
The dashed line represents the limit for thermally unstable mass transfer. 
\textit{Bottom}: Comparison between the observed (dark gray) and simulated
(light gray) WD mass distribution. Apparently, both models can not reproduce the observations. 
}
\label{fold}
\end{figure*}

Using the fitting formula for the secondaries
Roche-lobe provided by \citet{eggleton83-1},
Eq.\,(\ref{eq-f}) can be converted to: 
\begin{equation}
\zeta_{ad}=\frac{2}{3}\frac{ln(1+q^{1/3})-\frac{1}{2}\frac{q^{1/3}}{(1+q^{1/3})}}{0.6q^{2/3}+ln(1+q^{1/3})}(1+q)+2(q-1)
\end{equation} 
which can be solved to obtain the critical mass ratio above 
which mass transfer is dynamically unstable.
For the fully conservative case we obtain
$q_{\mathrm{crit}}=0.634$ for $\zeta_{ad}=-1/3$ 
and a steep increase in the range $M_2=0.5-0.7\Msun$. 
The results of the corresponding BPS model are 
illustrated in Fig.\,\ref{fold} (left panels).
We have plotted the predicted 
CV population in a mass ratio ($q=M_2/M_1$) 
versus secondary mass ($M_2$) diagram (top panel)
where the regions forbidden either due to 
dynamically unstable mass transfer or the 
WD mass exceeding the Chandrasekhar limit 
are gray shaded. The predicted current 
CV population is represented by the cyan dots 
while the black squares indicate the 
position of observed CVs with relatively robust 
measurements of both stellar masses.
The observed sample consists of the fiducial 
systems listed in ZSG11
and three more recently identified systems with 
robust mass estimates for both stars:
KISJ 1927+4447 \citep{littlefairetal14-1}, HS0220+0603 \citep{rodriguez-giletal15-1} and PHL 1445 \citep{mcallisteretal15-1}.

In the bottom left panel of Fig.\,\ref{fold} 
we compare the simulated (light gray) and 
observed WD mass distribution
and find the expected disagreement: 
the model predicts far too many 
CVs containing low-mass WDs. 

\subsection{The classical non-conservative model}

In more consistent evolutionary models of CVs, 
mass 
loss due to nova eruptions is taken into account.
In binary evolution models this can be incorporated 
as a continuous effect as shown by 
\citet{schenkeretal98-1}. 
Usually it is assumed that the specific 
angular momentum of the ejected mass equals
the specific angular momentum of the 
WD. This has a stabilizing effect on the 
mass transfer.
The corresponding CAML is
\begin{equation}
\frac{\dot{J}_{\mathrm{CAML}}}{J}=\nu\frac{\dot{M_2}}{M_2}
\end{equation}
with $\nu=M_2^2/(M_1(M_1+M_2))$
\citep[see e.g.][]{king+kolb95-1}. 
This CAML has to be taken into account when 
evaluating 
the right hand side of Eq.(\ref{eq-f}). 
Compared to the fully conservative 
case, CAML brings the system closer to 
dynamically 
unstable mass transfer as mass transfer 
generates 
additional angular momentum loss and reduces 
the Roche-volume of the secondary. 
However, in the classical CAML case 
this effect is more than 
compensated by the mass loss of the primary 
which significantly reduces the decrease of the 
secondaries Roche-lobe. 
The critical mass ratio is again defined by requesting
$\zeta_{\mathrm{ad}}=\zeta_{\mathrm{R_{L2}}}$ which results in:
\begin{equation}\label{eq-cCAML}
\zeta_{ad}=\frac{2}{3}\left(\frac{ln(1+q^{1/3})-\frac{1}{2}\frac{q^{1/3}}{(1+q^{1/3})}}{0.6q^{2/3}+ln(1+q^{1/3})}\right)+2\nu+\frac{M_2}{(M_2+M_1)}-2.
\end{equation}
We performed BPS calculations 
using the stability limit imposed by Eq.\,(\ref{eq-cCAML}) 
and taking into account the corresponding mass and  
angular momentum loss. The results are 
shown in the right panel of 
Fig.\,\ref{fold}. The disagreement between 
observations and theoretical prediction is even 
worse 
than in the fully 
conservative case (left panel), because even more CVs with low-mass WD and low-mass secondary 
star ($\lesssim\,0.35\Msun$)
are predicted to exist. 
Such systems are not found in the observed sample. 
As outlined in the introduction, neither 
observational biases nor our ignorance 
of the details of CE evolution, nor 
WD mass growth during or prior to CV evolution 
can solve this WD mass problem
(see ZSG11 and \citealt{wijnenetal15-1} for details). 

Given the large uncertainties affecting the limits 
of dynamically unstable mass transfer in models of
CV evolution and benefiting from the improved 
observational data,
we present in the next section an empirical model for 
CAML aiming to reproduce the observed WD masses. 

\subsection{An empirical CAML model}

\begin{figure*}
\includegraphics[width=0.73\textwidth]{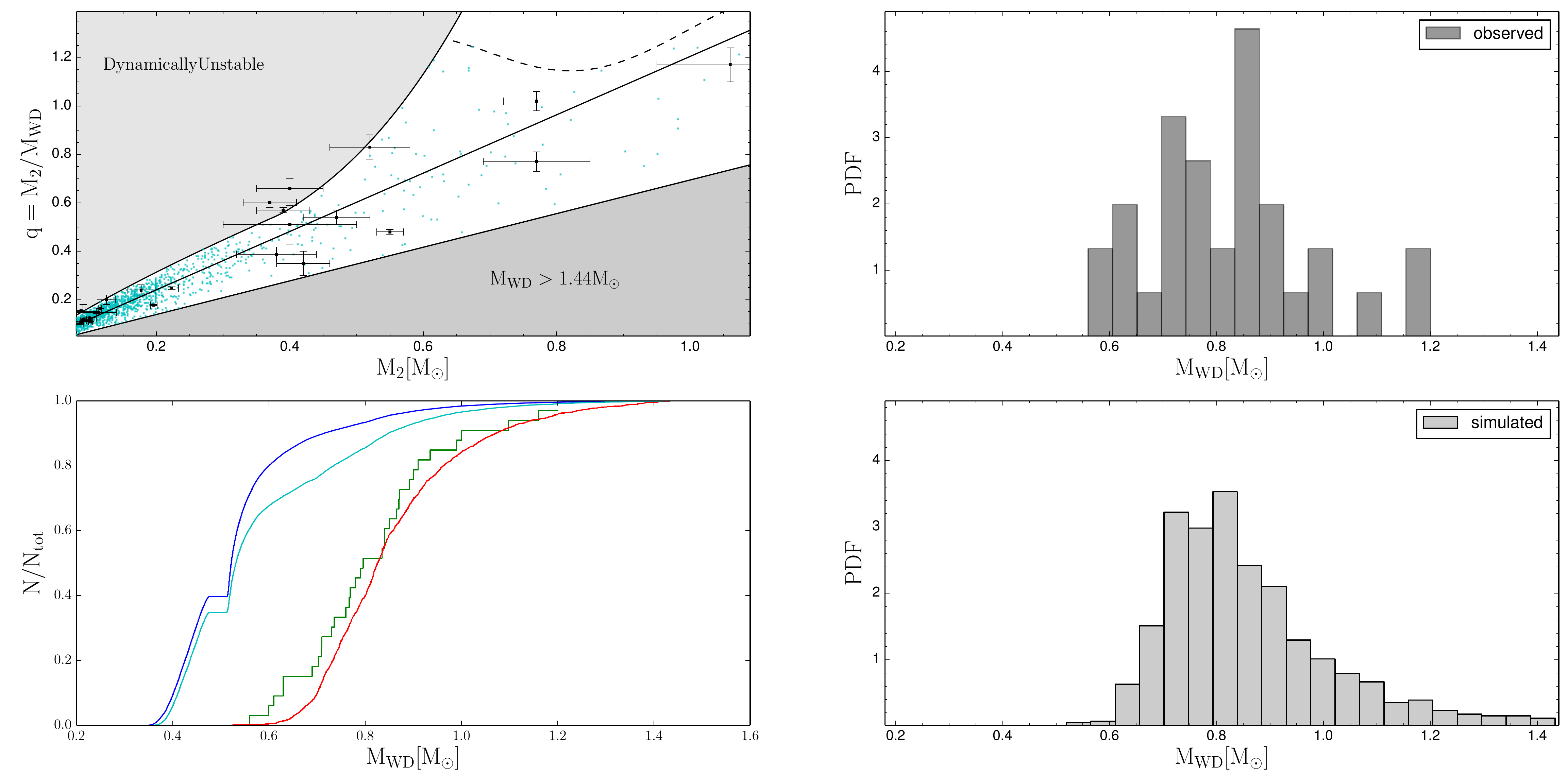}
\caption{\textit{Top left}: Observed and predicted
CV populations in the q versus $M_2$ diagram 
for our empirical CAML model. Colours and symbols
are the same as in Fig.\,\ref{fold}.
\textit{Right}: Observed (\textit{top}) and simulated 
(\textit{bottom}) WD mass distribution for the empirical 
CAML model.
\textit{Bottom left}: Cumulative distribution
of WD masses in CVs for the observed systems (green)
and for our three models: 
fully conservative (cyan), 
classical non-conservative (blue) and empirical 
CAML (red). Clearly, the latter model can 
reproduce 
the observations while the predictions of the other two models 
dramatically disagree with the observations.}
\label{fnew}
\end{figure*}

To investigate if and which form of CAML 
could solve the WD mass problem, we 
keep the standard assumptions used in 
the previous sections
but interpret the angular momentum loss 
associated with mass loss as a free parameter. 
Given that the disagreement between 
the observed and predicted WD masses in CVs 
is mainly caused by the large number of predicted
CVs with low-mass WDs below the period gap
(see Fig.\,\ref{fold}), we expect significant 
improvement if $\dot{J}_{\mathrm{CAML}}$ is stronger 
for low-mass WDs. 

Simulating CV populations for different 
forms of CAML and comparing with the 
observations we find good agreement if we assume:
\begin{equation}
\nu(M_1)=\frac{C}{M_1}
\end{equation}
with $C=0.3-0.4$. In the top left panel of 
Fig.\,\ref{fnew} we show 
the observed and predicted CV populations in the $q$ 
versus $M_2$ diagram for a simulated population of 
CVs assuming this empirical CAML (eCAML from now on)
model using $C=0.35$. The right panel shows the observed (top) and 
predicted (bottom) WD mass distribution 
while the bottom left panel shows 
the cumulative distribution of WD masses in CVs 
for the simulated samples 
(blue, cyan and red for the
fully conservative, classical 
non-conservative 
and eCAML model, respectively) 
and for the observed systems (green). 

Not too surprisingly, 
adding a new free parameter 
can solve the WD mass problem. 
However, before the eCAML model can be considered 
a viable option for CV evolution, 
we need to investigate how its predictions compare 
with other observed properties of CVs and we need to 
find a physical mechanism that might be 
responsible for eCAML. 
%
%
\section{Orbital period distribution 
and space density}

As described in the introduction, previous 
binary population models of CVs failed 
to reproduce both the predicted space density of CVs 
and the orbital period distribution. 
We now show that incorporating eCAML not only 
solves the WD mass problem but also brings into 
agreement the predicted and observed space density 
and orbital period distribution of CVs. 

\begin{figure}
\begin{center}
\includegraphics[angle=270, width=0.45\textwidth]{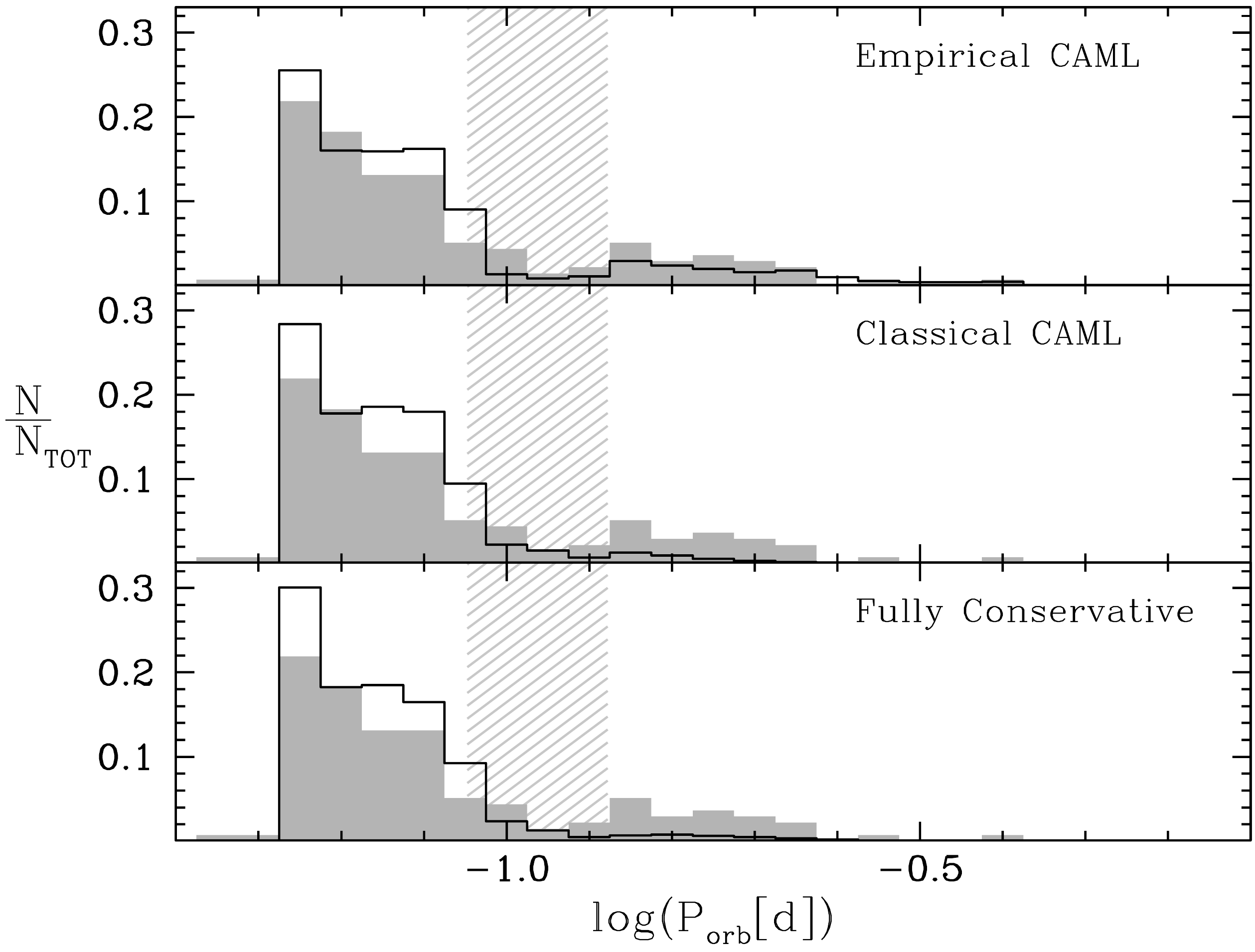}
\caption{Orbital-period distributions for our three 
models of CV population: 
eCAML (\textit{top}), classical CAML (\textit{middle}), 
and the fully conservative (\textit{bottom}). In grey the observed orbital
period as measured by \citet{gaensickeetal09-2}. The eCAML model is not only
the first model able to reproduce the observed WD mass distribution 
but also the first one to be in agreement with the 
observed orbital period distribution. }
\label{fporb}
\end{center}
\end{figure}

The orbital period distribution of
all discovered CVs with measured orbital period 
is heavily biased because CVs with more massive 
secondaries, i.e. those with longer orbital periods, 
are significantly brighter than short orbital 
period CVs. The sample of CVs least affected 
by this bias has been presented 
by \citet[][]{gaensickeetal09-2}, thanks to the 
depth of SDSS. In Fig.\,\ref{fporb} we compare 
the orbital period distribution of SDSS CVs 
with those predicted by the three models 
discussed in this paper. 
The classical CAML and the fully
conservative model do not only dramatically disagree 
with the WD mass distribution but are also 
unable to reproduce the observed period 
distribution
predicting the existence of too many CVs 
at short orbital periods and too few
above. In contrast, the eCAML model that was 
designed to solve the WD mass problem agrees 
very well with the observed period distribution. 
The fraction of systems predicted above the gap 
(15 per cent above 3.18 h)
is only slightly smaller than the observed 
fraction (19 per cent). 
 
The period distribution problem is not the only 
long standing issue the eCAML approach
solves simultaneously with the WD mass problem. 
As a large number of systems containing 
low-mass WDs is supposed to suffer from unstable 
mass transfer, the total number of CVs 
predicted by the eCAML model is 
smaller by a factor of $\sim0.11$ and $\sim0.24$ 
compared to the classical CAML and the fully 
conservative model, respectively, which should
bring the predicted CV space density into the
range of the values derived from observations. 

\section{Physical interpretation of eCAML}

The eCAML model presented in this paper can 
simultaneously solve the three biggest problems 
of CV evolution. However, there must be 
a physical mechanism behind this parameterised 
model. 

The most obvious mechanism that may generate 
significant CAML in CVs are nova eruptions. 
In fact, the classical CAML model represents
a weak form of CAML caused by novae.
This model assumes the 
expelled material to take away the specific 
angular momentum of the WD which represents a reasonable assumption
\emph{if} friction between the ejecta and the 
secondary star is negligible. If, 
on the other hand, friction contributes
significantly, CAML might be much stronger 
than predicted by the classical CAML model. 

According to a detailed study of 
\citet{schenkeretal98-1}, 
frictional angular momentum loss produced 
by novae only weakly depends on the 
mass ratio but is very sensitive to the 
expansion velocity of the ejecta, 
i.e. assuming slow expansion velocities,
the consequential (frictional) angular 
momentum loss can be significant
\citep[see][their Figure 5]{schenkeretal98-1}.  
Interestingly, nova models predict significantly 
faster expansion velocities for CVs containing 
high-mass WDs than for those containing low-mass 
WDs \citep[see e.g.][their Fig.\,2c]{yaronetal05-1}. 
The CAML generated by slow nova eruptions in CVs 
containing low-mass WDs might therefore 
be the physical cause behind the 
eCAML model we used to bring into 
agreement observations and theory.

To test this hypothesis, the velocity of nova 
ejecta at the position of the 
secondary as a function of WD 
mass needs to be determined and
incorporated in detailed models 
of frictional angular momentum loss.  
Observationally, this would require to perform 
spectroscopy at the onset and throughout a 
nova which is probably unrealistic 
as currently the occurrence of nova
eruptions can not be predicted. Thus, testing 
the eCAML hypothesis will probably rely 
on improving theoretical models of novae 
for some time to come.  

\section{Conclusion}

Since the work of \citet{zorotovicetal11-1}, the 
WD mass problem has become the most serious 
limitation of our understanding of CV evolution.
In \citet{wijnenetal15-1} we have shown that neither 
assuming WDs in CVs to
grow in mass nor assuming a preceding phase of 
thermal time scale mass
transfer for many CVs can solve the problem without 
violating other observational facts. 
Here we have presented a new model for CV evolution 
assuming 
consequential angular momentum loss (CAML) 
that increases with decreasing WD
mass. In this model the previously predicted CVs 
with low mass WDs at short orbital periods suffer 
from dynamically unstable mass transfer and will
disappear as CVs. This empirical CAML (eCAML) 
model is currently the only available model 
that can reproduce the observed WD mass 
distribution. 
Simultaneously and without any (further) 
fine-tuning, the model also solves the long standing 
disagreement between the predicted and observed 
orbital period distribution and space
density of CVs. This makes eCAML a promising 
candidate for being the missing piece 
in the CV evolution puzzle. 
The best candidate for the physical mechanism 
behind eCAML is frictional
angular momentum loss following nova eruptions
because CVs with low-mass WDs produce slower novae which 
might lead to increased frictional angular momentum loss. 
To test this hypothesis realistic models for nova 
eruptions and the resulting frictional 
angular momentum loss need to be developed.

\section*{Acknowledgments}

MRS thanks Gijs Nelemans and Boris Gaensicke 
for interesting and helpful discussions
at the \emph{Physics of Cataclysmic
and Compact Binaries} workshop held in NY, 
November 2014.  
MRS and MZ thank for support from Fondecyt 
(1141269 and 3130559) and Millennium Nucleus RC130007
(Chilean Ministry of Economy). 
TPGW is grateful for funding support by the Netherlands Organisation for Scientific Research (NWO) under grant 614.001.202.





\bibliographystyle{mn2e}





\bsp	
\label{lastpage}
\end{document}